# USING SOHN'S LAW OF ADDITIVE REACTION TIMES FOR MODELING A MULTIPARTICLE REACTOR. THE CASE OF THE MOVING BED FURNACE CONVERTING URANIUM TRIOXIDE INTO TETRAFLUORIDE

F. Patisson, B. Dussoubs, and D. Ablitzer
Laboratoire de Science et Génie des Matériaux et de Métallurgie (LSG2M)
UMR 7584 CNRS-INPL, Nancy School of Mines
Parc de Saurupt, 54042 Nancy Cedex, France

## Abstract

One of the major issues with multiparticle reactors is to handle their multiscale aspect. For modeling, it usually comes to coupling a reactor model (describing the phenomena at the macroscopic scale) with a so-called grain model (simulating the behavior of a single grain or a particle). An interesting approach proposed by H.Y. Sohn (1978) is to use the law of additive reaction times in order to calculate, approximately but analytically, the reaction rate of a particle in the reactor model. Its great advantage, compared to a numerical grain model, is to drastically reduce the computation time, particularly in the case of complex reactor models. This is the approach we retained for modeling the moving bed furnace, a counter-current gas-solid reactor used in the nuclear fuel-making route for producing uranium tetrafluoride from uranium trioxide. The numerical model we developed is 2-dimensional, steady-state and based on the finite volume method. It describes solid and gas flow, convective, conductive and radiative heat transfers, and six chemical reactions involved in the process. The law of additive reaction times is used to calculate analytically the rate of the three principal gas-solid reactions at every discrete point in the reactor. We have demonstrated the validity of this approach by comparing its results with those calculated from a numerical grain model. Also detailed in the paper are the main results of the moving bed furnace model itself and the possibilities of optimizing the process revealed by the calculations.

## Introduction

Gas-solid reactions are at the root of numerous major industrial processes. Notable examples are the combustion of solids, the incineration of wastes, the absorption of acid gases and reactive vapor phase deposition. In the field of pyrometallurgy one can cite the treatment of ores by calcination, roasting and reduction, the oxidation of metals, the pyrolysis of coal, the oxidation of coke. In most cases the kinetic and thermal characteristics of the reaction(s) control the plant productivity and impose their requirements for operating the process (range of temperatures to abide by, profile of flow rate and gas composition to control, etc.). This explains the great interest for the knowledge and the control of these reactions and their kinetics, and consequently, the numerous studies devoted to the mathematical modeling of both gas-solid reactions and gas-solid reactors.

The reactors involved are of all types, *e.g.* bed of particles traveling on a moving grate, moving beds in shaft furnaces, rotary kilns, fluidized beds, batch fixed beds. Features shared by most of these reactors are, in addition to the presence of gas-solid reactions, a high-temperature operation

(700-1300 °C), the treatment of divided solids, often heterogeneous and poorly defined, and phenomena occurring at several scales, from meter to micrometer. For mathematical modeling, the latter feature is quite significant since it drives the choice of the modeling approach, as it will be discussed in detail in the present paper. Roughly speaking, one has to combine a description at the scale of the solid particles, the size of which ranges from centimeter (pellets, lumps) to micrometer (powders) and which often include micrometric pores, with a description at the scale of the reactor, typically a few meters.

## The Single Particle

<u>Main Physical Phenomena</u>

The solid particles treated industrially are generally porous pellets composed of a number of individual grains of different sizes, separated by the pores. If these grains are themselves porous, they can be considered as a combination of dense crystallites and pores. A possible representation of the solid is then that shown in Figure 1a, where the pellet is in fact an agglomerate of dense grains of different sizes. Let us consider a pellet of this type immersed in a gas, and inside which the following reaction occurs

$$aA_{(g)} + bB_{(s)} = pP_{(g)} + qQ_{(s)} \tag{1}$$

($a$, $b$, $p$, and $q$ are the stoichiometric coefficients, and $A$, $B$, $P$, and $Q$ the species), and examine in more detail the physical, chemical and thermal phenomena that take place.

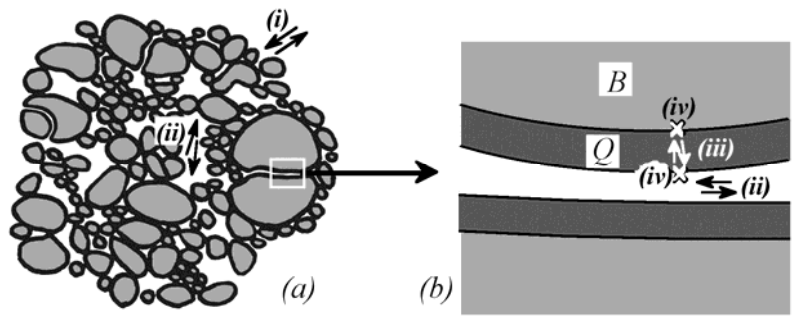

Figure 1. Schematic representation of a porous pellet (*a*) made up of grains, a pore (*b*), and the steps in mass transport (*i* to *iii*) and reaction (*iv*). B and Q are the solid reactant and product phases.

Before reacting with the solid *B*, the reactive gas *A* must be transported to the reaction site by *(i)* transfer within the external gas to the pellet surface and *(ii)* diffusion through the intergranular pores. In the grains (Figure 1b), the reaction *(iv)* can itself be broken down into adsorption, reaction, and desorption steps, and generally also involves a final mass transport step *(iii)*, through the layer of solid product *Q*, which can be either dense or porous. The gaseous product *P* is desorbed, then evacuated outwards through the pores. Each of these steps has its own kinetics and can limit, or help to limit, the overall rate of conversion. A so-called chemical regime is one where the chemical reaction itself controls the overall kinetics, while a diffusional regime is one governed by diffusion of the gas either in the pores or in the layer of *Q*. Other possibilities are mixed regimes, corresponding to various combinations of the controlling steps.

If the reaction is exo- or endothermic, or if the external temperature varies for any reason, the evolution or consumption of heat by the reaction, and heat transfer within the pellet and to the outside are superimposed on the mass transport processes. In the case of thermal decomposition reactions (endothermic, with *a*=0, e.g. the decomposition of calcium carbonate), the heat transfer can itself control the overall kinetics.

Finally, a characteristic feature of numerous gas-solid reactions is the variation of the porous structure of the solid in the course of reaction. This evolution can be caused by external stresses, by high-temperature sintering or simply by the chemical reaction itself. The conversion of the solid phase *B* to a new solid phase *Q* of different molar volume modifies the morphologies of the solid grains, the pores and sometimes the pellet.

Grain, Pore and Pseudo-Homogeneous Models

The large variety of physical phenomena involved explains why it is difficult to describe mathematically the transformation processes in all their complexity. Nevertheless, an abundant literature deals with the topic and, since the early Shrinking Core Model (SCM) [1], mathematical models aiming at simulating gas-solid reactions in a single particle became increasingly sophisticated. A possible way to classify these models is to follow the underlying description of the particle structure (Figure 2). Grain models image the structure depicted in figures 1a and 2a, *i.e.* a pellet made up of grains. The pioneering work of Szekely, Evans and Sohn [2] was later extended to account for a variable porosity [3] and changing and distributed grain sizes [4, 5]. Grain models can account for the chemical reaction itself, together with the associated mass transport phenomena, like the diffusion through the solid product layer, the diffusion through the intergranular pores, and the external transfer. As they represent a significant improvement on earlier, much simpler models, grain models were widely and successfully applied to the simulation of numerous gas-solid reactions. The second class, pore models, first considers the pores, the solid phase being the volume complement of the pore phase [6-10]. These are well adapted to the description of the pore evolution in the course of the reaction [11, 12]. Eventually, all these models mostly reflect the same physical and chemical phenomena and the description of the structure in terms of grains or pores leads essentially to different expressions for the evolution of the reaction surface area [13, 14]. This justifies the development of another kind of models, the pseudo-homogeneous ones, which do not describe a grain or pore structure *a priori*, but only through the expressions of the reaction rate and the effective transport properties [14, 15].

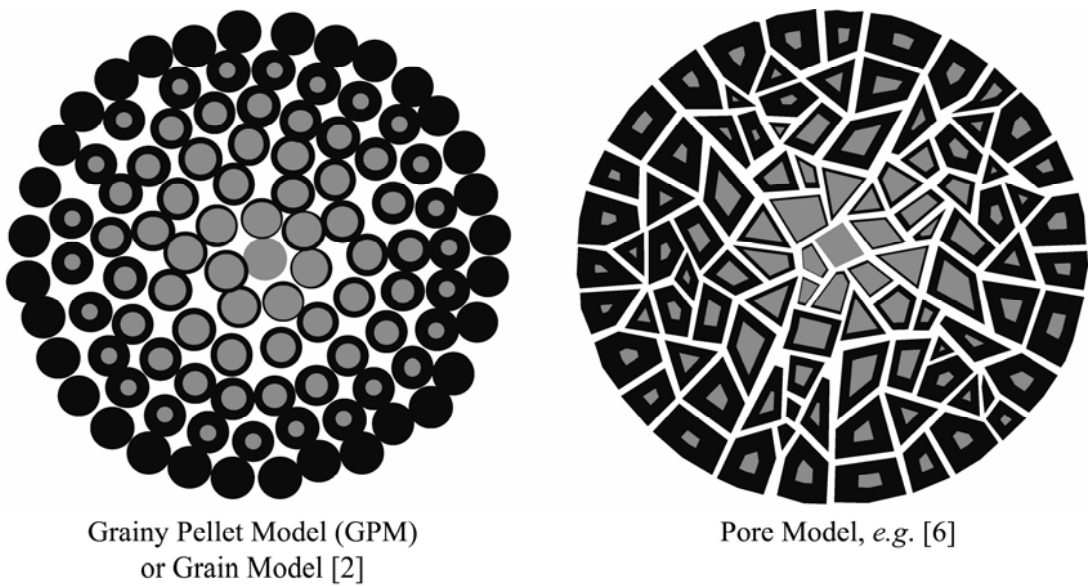

Figure 2. Grain and pore models

Whatever the type, most of the models require a numerical solution. Only simple situations and rate expressions can be treated analytically. At least two coupled equations are to be solved, these expressing the local balances of the gaseous and solid reactant. Table I gives the equations solved in the pseudo-homogeneous model "Boulet" [14]. The latter was developed by two of the authors in order to simulate both the kinetic and the thermal behavior of a porous spherical pellet in reaction with a gas. Non-isothermal and non-equimolar reactions can be dealt with, possibly in the transient regime and in the presence of gaseous and solid inerts. A subsequent version was written accounting for the evolution of a pore size distribution [16]. This model has been validated against analytical and numerical solutions from the literature and was successfully applied for simulating the oxidation of zinc sulfide and the hydrofluorination of uranium dioxide.

Table I. Main equations of the model "Boulet" [14].

| Equation | Description |
|---|---|
| $\text{div}\,\vec{N}_A + \frac{\partial}{\partial t}(\varepsilon c_t x_A) = -a\,r_V$ | Balance in $A$ |
| $\text{div}\,\vec{N}_P + \frac{\partial}{\partial t}(\varepsilon c_t x_P) = p\,r_V$ | Balance in $P$ |
| $\text{div}\,\vec{N}_t + \frac{\partial}{\partial t}(\varepsilon c_t) = (p - a)r_V$ | Total gas balance |
| $c_{B_0} \frac{\partial x_B}{\partial t} = b\,r_V$ | Balance in $B$ |
| $c_{pV} \frac{\partial T}{\partial t} + \text{div}\left(-\lambda_e \overrightarrow{\text{grad}\,T}\right) = r_V(-\Delta_r H)$ | Heat balance |
| $\vec{N}_A = x_A \vec{N}_t - D_{Ae} c_t \overrightarrow{\text{grad}\,x_A}$ | Flux density of $A$ |
| $\vec{N}_P = x_P \vec{N}_t - D_{Pe} c_t \overrightarrow{\text{grad}\,x_P}$ | Flux density of $P$ |
| $\vec{N}_t = \vec{N}_A + \vec{N}_P + \vec{N}_I$ | Total flux density |
| $r_s = k_0 e^{-\frac{E_a}{RT}} \left( c_{A_s}^n - \frac{c_{P_s}^l}{K_{eq}} \right)$ | Surface rate |
| $r_V = r_s\, a_0\, s(x_B)$ | Volume rate |
| $s(x_B) = (1 - x_B)^m$ | Evolution of the reaction surface area |
| $X_B = \frac{1}{V_p} \int_{V_p} x_B\, dV_p$ | Overall conversion |

Notation: $N_A$, $N_P$, $N_I$, $N_t$, molar flux densities of $A$, $P$, $I$ (inert), and total; $t$ time; $\varepsilon$, porosity; $c_t$, total gas concentration; $x_A$, $x_P$, molar fractions of $A$, $P$; $r_V$, reaction rate per unit volume; $c_{B_0}$, initial molar concentration in $B$; $x_B$, local conversion; $c_{pV}$, volumic specific heat; $\lambda_e$, effective thermal conductivity; $\Delta_r H$ heat of reaction; $D_{Ae}$, $D_{Pe}$, effective diffusivities; $r_s$, reaction rate per unit surface area; $E_a$, energy of activation; $n$, $l$, and $m$, reaction orders; $K_{eq}$, equilibrium constant; $a_0$, initial specific surface area; $V_p$ particle volume.

The Law of Additive Reaction Times

To the authors' knowledge, the first mentions of the law of additive reaction times are found in [2] when Szekely, Evans and Sohn evaluate the different mass transfer resistances in the case of a reaction in the mixed regime (reaction-gas diffusion) in the frame of the Grain Model. Sohn little later stated more clearly this law under the following form [17]:

$$\begin{pmatrix} \text{time required to attain} \\ \text{a certain conversion} \end{pmatrix} = \begin{pmatrix} \text{time required to attain} \\ \text{the same conversion} \\ \text{in the absence of resistance} \\ \text{due to intrapellet diffusion} \\ \text{of fluid reactant} \end{pmatrix} + \begin{pmatrix} \text{time required to attain} \\ \text{the same conversion} \\ \text{under the control of} \\ \text{intrapellet diffusion} \\ \text{of fluid reactant} \end{pmatrix} \quad (2)$$

Mathematically, and adding the possible contribution of external transfer, it can be written:

$$t_{X_B} = t_{X_B,\text{chemical regime}} + t_{X_B,\text{diffusion regime}} + t_{X_B,\text{external transfer regime}} \quad (3)$$

This law is exact in the case of a non-porous particle following the SCM with a reaction order with respect to the gas reactant equal to 1. Szekely *et al.* and Sohn, however, had the intuition and showed [17] that it holds, approximately, in a number of situations: these described by the original Grain model, for different pellet and grain shapes, but also the case when diffusion through the product layer around grains is a rate-limiting step, as well as for various reaction kinetics, like those of the nucleation-and-growth type or first-order with respect to the unreacted solid. The physical reason underlying the additivity of the reaction times is that these times are tantamount to mass transfer resistances. The resistances are perfectly in series for a non-porous pellet, and roughly in series for a porous one.

The calculation of the reaction times is problem-specific. According to the Grain Model and for spherical pellet and grains, one has:

$$t_{X_B} = \tau_{ch}\left[1-(1-X_B)^{\frac{1}{3}}\right] + \tau_{dif}\left[1-3(1-X_B)^{\frac{2}{3}}+2(1-X_B)\right] + \tau_{ext} X_B$$

$$\text{with} \quad \tau_{ch} = \frac{3c_{B_0}}{ba_0 k_r c_{A\infty}} \quad , \quad \tau_{dif} = \frac{3ac_{B_0}}{2ba_{ext}^2 D_{eA} c_{A\infty}} \quad , \quad \tau_{ext} = \frac{ac_{B_0}}{ba_{ext} k_g c_{A\infty}} \quad (4)$$

Our own notations are used here. $\tau_{ch}$ is the characteristic time of the chemical reaction itself, i.e. the time that would be necessary to get a complete conversion under chemical reaction control, $\tau_{dif}$ the time for complete conversion under gas diffusion control, and $\tau_{ext}$ the time for complete conversion under control by the external transfer. In addition to the notations already used, $a_{ext}$ is the external surface area of the pellet, $c_{a\infty}$ the molar concentration in $A$ in the bulk external gas, and $k_g$ the external mass transfer coefficient.

As indicated by Sohn, a major benefit of the law of additive reaction times is the possibilities offered by its analytical, "closed-form" expression. The calculation of the evolution of the conversion as a function of time is hundreds of times faster using (3) than using a numerical grain model, which implies a time-consuming spatial discretization of the pellet. This is a decisive advantage when attempting to model and simulate multiparticle reactors. In this case, the proper variable to handle is the reaction rate, which can be obtained by the derivation of (3) – or (4) for the Grain Model. In the latter case, one obtains:

$$r_{X_B} = \left[\frac{\tau_{ch}}{3}(1-X_B)^{-\frac{2}{3}} + 2\tau_{dif}\left[(1-X_B)^{-\frac{1}{3}}-1\right] + \tau_{ext}\right]^{-1} \quad (5)$$

where $r_{X_B}$ is the reaction rate expressed as the time derivative of the conversion. Sohn tested and checked against a numerical grain model the validity of using the derivative form of the law of additive reaction time when temperature and gas composition change with time [17].

**The Multiparticle Reactor**

The purpose of this section is not to present all of the approaches for modeling multiparticle processes. We rather intend to show how gas-solid reactions can be described and treated mathematically in multiparticle reactor models. The most reliable mathematical models of processes are certainly those describing the actual transport phenomena involved and using the formalism of Computational Fluid Dynamics (CFD), *i.e.* based on the solution of the local balance equations. Compared to the case of a single particle, the physical-chemical phenomena to account for at the multiparticle reactor scale are the same, with the addition of gas and solid flows. This kind of model requires a numerical solution. The reactor is meshed in 2-D or 3-D and treated either in steady or transient state. The usual current numerical methods are finite elements or finite volumes and the codes are either multi-purpose commercial CFD softwares or specifically-written, dedicated ones.

Coupling Particle and Reactor Models

The issue of modeling a multiparticle reactor in which gas-solid reactions take place is clearly a multi-scale problem. As stated before, tools are available for describing finely the phenomena at the scale of the particles (from µm to mm), these are the particle or grain models, and others, the commercial or dedicated softwares, based on balance equations on elementary reactor volumes, are available to deal with the macroscopic scale (roughly from mm to m). Coupling, i.e. making these two types of numerical codes enter into dialog or being totally integrated, raise the theoretical question of the representative elementary volume and the practical one of the coupling method.

At the macroscopic scale of the process, it is desirable and convenient to be able to model the two-phase (gas+solid) content of the reactor as a continuous and homogeneous medium. This wish, common to most of scientific domains dealing with heterogeneous media, gave birth to the concept of Representative Elementary Volume (REV). The latter can qualitatively be defined as a volume large enough for defining meaningful "equivalent" properties (equivalent to those of a continuous homogeneous medium) but small enough compared to the macroscopic dimensions. A lot of work was devoted to the calculation of these equivalent properties, like the effective thermal conductivity, the effective mass diffusivities, or the permeability, in particular by means of local averaging techniques on the REV [18]. We assume here that this approach is valid for gas-solid reactors.

Concerning the method of coupling, the simplest one is to integrate the particle model into the reactor model, which, in terms of programming, amounts to making the particle model a subroutine of the reactor model. Coupling is carried out through the reaction rate. The reaction rate per unit reactor volume, $r_{V_r}$, which appears in nearly all balance equations, is expressed as:

$$r_{V_r} = \frac{1}{b} c_{B_{0r}} r_{X_B} \qquad (6)$$

where $c_{B_{0r}}$ is the initial apparent concentration in *B* per unit reactor volume. $r_{X_B}$ is given by the particle model as a function of the particle state (internal profiles of $x_B$, $x_A$, $x_P$, $T$, etc.) at two consecutive times and of the local conditions prevailing around the particle (essentially the temperature, composition, and velocity of the gas). To calculate correctly $r_{X_B}$, it is thus necessary, for every reactor cell, to store two complete states of the particle and to run the particle model, which implies a spatial integration along the particle radius (for a 1D particle model). Such a method of coupling thus requires, compared to the calculations that would be necessary for a homogeneous reaction, one additional level of integration, equivalent to a fourth dimension. For reactors of simple geometry (1D or 2D) and in steady state, this approach remains numerically practicable. An example treated this way by two of the authors is the modeling of rotary kilns for coal pyrolysis [19, 20]. For finer descriptions (2D or 3D) at the scale of the reactor, the computation time may become unacceptable.

Using the Law of Additive Reaction Times

In the latter situation, the interest of a simpler, though accurate, means for calculating the reaction rate is clear. For every reactor point, the reaction rate is no longer computed numerically by a particle model, but given by an equation similar to (5). It suffices to calculate the characteristic times as a function of the local gas and temperature conditions. These times can even be calculated in cases that go beyond the rather restrictive frame of the Grain Model. Other systems than a particle, like beds of powder in a lifter or at the bottom of a rotary kiln, can also be described considering additive reaction times [21].

Surprisingly, despite the advantages cited above, very few papers mentioning the use of the law of additive reaction times for modeling multiparticle processes were published. In addition to the present work, other examples are referred to in [17]. A recent approach using another approximate method is proposed by Gómez-Barea and Ollero, who note that this kind of approach has been far more frequently used in the field of catalytic reactions, but who seem to ignore the law of additive reaction times [13].

## The Moving Bed Furnace for the Production of UF$_4$

The Process

One of the numerous steps in the route for making the nuclear fuel, which comes after the purification and before the isotopic enrichment, is the conversion of uranium trioxide UO$_3$ into uranium tetrafluoride UF$_4$. In the French process, it is carried out in a moving bed furnace (Figure 3). UO$_3$ pellets (25-mm diameter) fed at the top of a long (9 m) cylinder, descend by gravity, and react with counter-flowing gases. In the upper sectio (the reduction zone) the reducing gas is hydrogen produced by cracked ammonia NH$_3$. In the lower section (the hydrofluorination zone) hydrogen fluoride is injected to convert UO$_2$ into UF$_4$. Particles of UF$_4$ are extracted in a horizontal section by an Archimedes screw. The tricky point for the process is its control. The reactions of reduction and hydrofluorination are very exothermic. Local hot spots can soften the pellets and lead to stopping of the bed. In order to build a numerical tool that could help to optimize the process and its operation, we developed a mathematical model dedicated to this process.

The Model

The problem is indeed rather complex. All types of transport phenomena take place and at least 5 heterogeneous and 1 homogeneous chemical reactions occur. Several pieces of work were necessary to complement the modeling task: establishing overall mass and heat balance of the industrial process from specific on-line measurements, design of a cold model to study solid and gas flow, separate studies of the cracking of ammonia, the reduction of UO$_3$ and of U$_3$O$_8$, the hydrofluorination of UO$_2$, etc. Table II summarizes the reactions and the kinetic laws retained here. Three of these (the third, fourth, and sixth equations) correspond to the law of additive reaction times. The characteristic times were not calculated as given by the standard Grain Model, but specifically to account for the characteristics of the reactions. For example, in the case of hydrofluorination, the non-equimolarity and the reversibility of the reaction are considered. The resulting expressions are reported below:

$$\tau_{ch,f} = \frac{2 d_g c_{UO_2,p}}{\left(1-\varepsilon_p\right) k_f \left(c_{HF} - \sqrt{c_{H_2O}/K_{eq}}\right)} \qquad (7)$$

$$\tau_{dif,f} = \frac{d_p^2 c_{UO_2,p}}{6 D_{eff} \left( c_{HF} - c_{HFeq} \right)} \frac{x_{HF} - x_{HFeq}}{2 \ln\left( \frac{2 - x_{HFeq}}{2 - x_{HF}} \right)} \quad (8)$$

$$\tau_{ext,f} = \frac{2 d_p c_{UO_2,p}}{3 k_g} \frac{1 - \frac{x_{HFeq}}{2}}{c_{HF} - c_{HFeq}} \quad (9)$$

where $f$ is a subscript for hydrofluorination, $d_g$ and $d_p$ are the diameters of the grain and the pellet, $c_{UO_2,p}$ the initial apparent molar concentration in $UO_2$ of the pellet, $\varepsilon_p$ the porosity of the pellet.

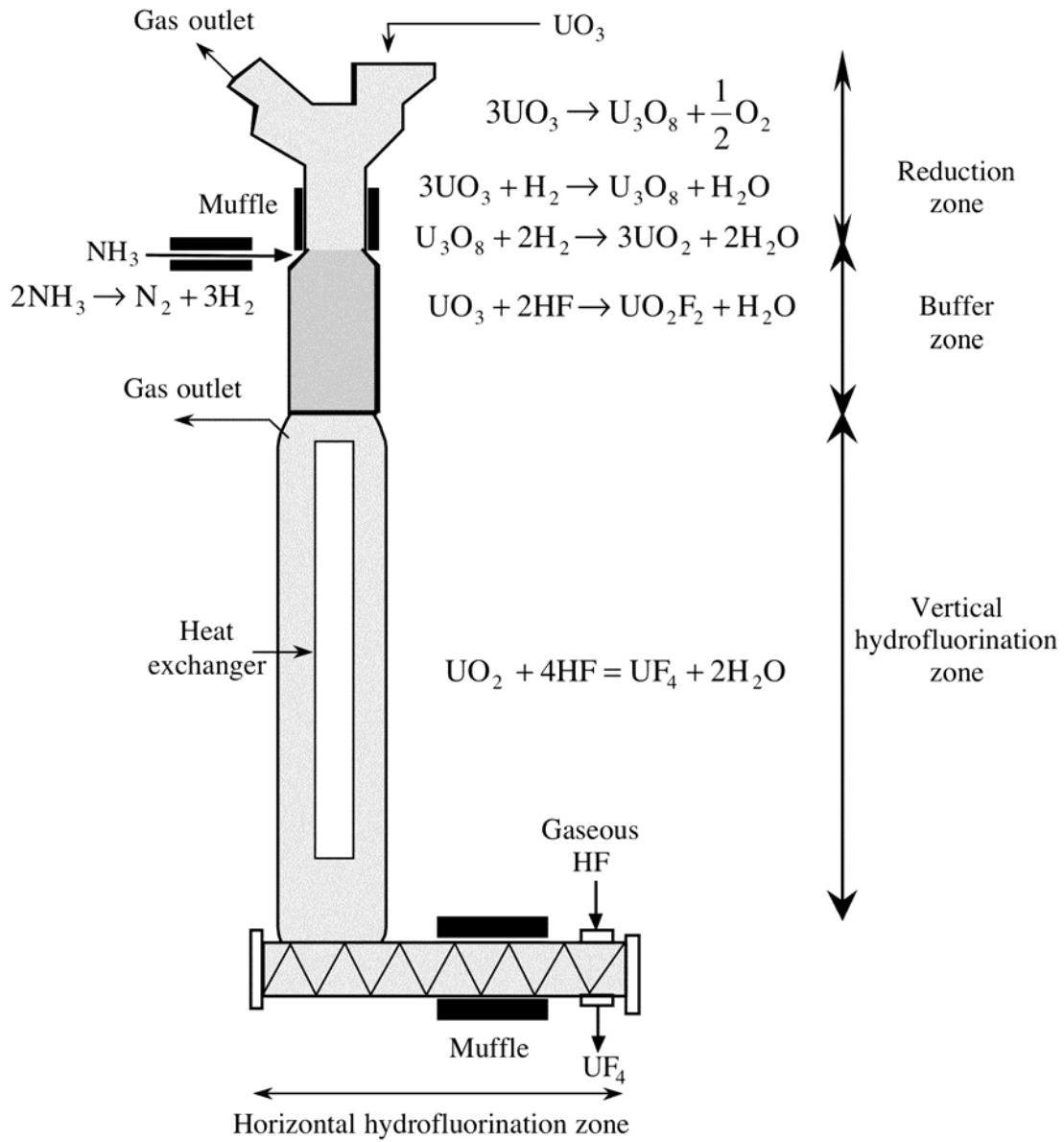

Figure 3. Schematic representation of the moving bed furnace producing $UF_4$.

Table II. The different chemical reactions considered and their rate expressions, from [22].

| | Reaction | Expression of the reaction rate /mol m$^{-3}$ s$^{-1}$ |
|---|---|---|
| **Reduction zone** | Cracking $\quad 2NH_3 \rightarrow N_2 + 3H_2$ | $r_{cr} = k_{cr} \exp\left(-\dfrac{E_{cr}}{RT}\right) c_{NH_3}$ |
| | Calcination $\quad 3UO_3 \rightarrow U_3O_8 + \dfrac{1}{2}O_2$ | $r_{cal} = k_{cal} \exp\left(-\dfrac{E_{cal}}{RT}\right) c_{UO_3}$ |
| | Reduction 1 $\quad 3UO_3 + H_2 \rightarrow U_3O_8 + H_2O$ | $r_{red1} = \dfrac{c_{UO_3,e}}{3}\left[\tau_{ch,red1} + 2\tau_{dif,red1}\left(\left(1-X_{UO_3}\right)^{-1/3}-1\right) + \tau_{ext,red1}\right]^{-1}$ |
| | Reduction 2 $\quad U_3O_8 + 2H_2 \rightarrow 3UO_2 + 2H_2O$ | $r_{red2} = c_{U_3O_8,e}\left[\dfrac{\tau_{ch,red2}(1-Z)}{3(1-Z^{1/3})}\left(1-(1-Z)X_{U_3O_8}\right)^{-2/3} + 2\tau_{dif,red2}\left(\left(1-X_{U_3O_8}\right)^{-1/3}-1\right) + \tau_{ext,red2}\right]^{-1}$ |
| | Secondary hydrofluorination $\quad UO_3 + 2HF \rightarrow UO_2F_2 + H_2O$ | $r_{hf_s} = k_{hf_s} \exp\left(-\dfrac{E_{hf_s}}{RT}\right) c_{UO_3} c_{HF}$ |
| **Hydrofluorination zone** | Hydrofluorination $\quad UO_2 + 4HF = UF_4 + 2H_2O$ | $r_f = c_{UO_2,e}\left[\dfrac{\tau_{ch,f}}{3}\left(1-X_{UO_2}\right)^{-2/3} + 2\tau_{dif,f}\left(\left(1-X_{UO_2}\right)^{-1/3}-1\right) + \tau_{ext,f}\right]^{-1}$ |

Concerning the model itself, its distinctive features are a multi-scale description, using the law of additive reaction times to model the main three heterogeneous reactions, and the combination of two modeling approaches: a mechanical one to describe the vertical sections, using the finite volume method in 2D, and a systemic one, using perfectly stirred reactors, to model the horizontal section with its Archimedes screw. For the model of the vertical section, gas and solid temperatures are assumed equal due to a high convective transfer between gas and solids. The model is further detailed in [22, 23].

The results calculated are in good agreement with available measurements, particularly the axial profiles of temperature recorded during a standard operation. From the calculated maps of temperature, compositions, reaction rates, and so on, some in-depth and new information about the way the furnace works could be obtained. For example, the existence, in the lower part of the hydrofluorination zone, of a large central hot region with little reaction taking place was revealed (see Figure 4). This is due to the presence of high radial temperature gradients and to a thermodynamic limitation at high temperature. This behavior was unknown by the operators and is eventually responsible for a poorer performance of the largest furnaces, as experienced industrially when comparing furnaces of different diameters. This also led us to propose possibilities of improvement of the process, *e.g.* by injecting a part of the hydrogen fluoride directly at the level of the hot region, the temperature would be decreased and the reaction would be favored. Indeed, these findings were applied to one of the industrial furnaces, specifically modified for that purpose, and operational results have confirmed the model predictions.

Comparison Between the Law of Additive Reaction Times and a Particle Model

To ensure the validity of using the law of additive reaction times, we have compared the conversion degree of a reacting UO$_2$ pellet first, when it was calculated with the model of the moving bed furnace and second, when it was calculated in the same conditions (temperature, gas composition) by means of a particle model. In the latter case, we used as a reference the model Boulet [14] presented above.

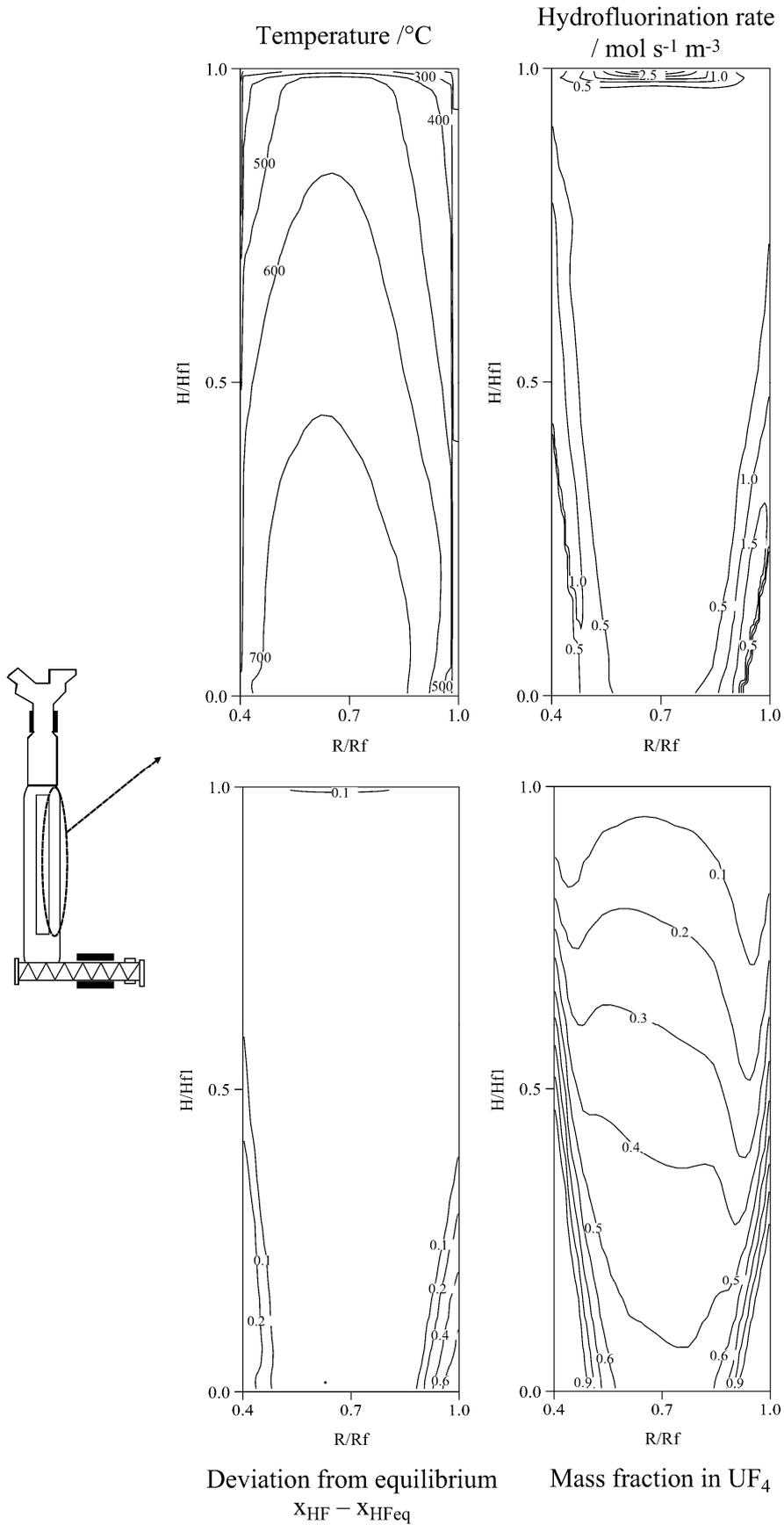

Figure 4. Calculated results regarding the vertical hydrofluorination zone.

Calculations were made for a pellet descending the upper part of the vertical hydrofluorination zone, with the parameters (thermodynamic and transport properties, diffusion coefficients, etc.) used in the model of the moving bed furnace [24]. Profiles of temperature and gas composition taken from the moving bed results were used as boundary conditions to the grain model. Two cases were considered: either the pellet descends in the centre of the moving bed (leading to the worst conversion into $UF_4$ at the bottom of the vertical zone), or next to the outer wall (thus leading to the best conversion).

Figures 5 and 6 present the mean conversion degree and the temperature of the centre of the pellet in these two cases. Figure 5 shows that the conversion of the pellet is very well represented with the law of additive characteristic times in the model of the moving bed. Figure 6 shows that, in spite of the exothermicity of the hydrofluorination, the gas temperature and the temperature at the centre of the pellet are very close, thus confirming the hypotheses of a uniform temperature in the pellet and of equal gas and solid temperatures.

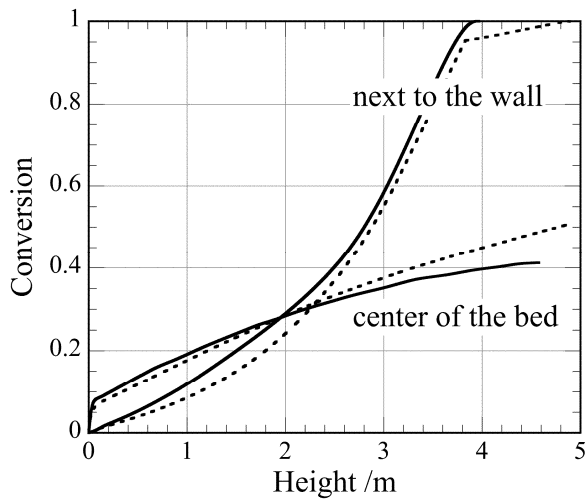

Figure 5. Evolution of the conversion of a pellet along the vertical hydrofluorination zone, for two radial positions. Solid lines: model Boulet [14], dotted lines: law of additive reaction times in the moving bed model, from [24].

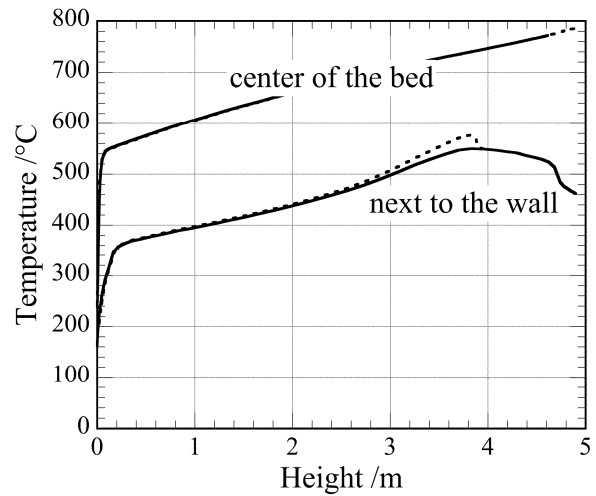

Figure 6. Evolution of temperatures along the vertical hydrofluorination zone, for two radial positions. Solid lines: at the center of a pellet, according to the model Boulet [14], dotted lines: gas temperature from the moving bed model, from [24].

Those results illustrate properly the validity of the hypotheses made in the complete model of the moving bed. They also represent, to our knowledge, the first numerical validation of using the law of additive characteristic times in a multiparticle reactor model.

## Conclusion

Gas-solid reactions associate chemical processes, for the chemical reaction itself, to physical mass transport processes for transferring the reactant and product species. We have briefly described these processes and the consequent kinetic regimes, together with the main mathematical models available from the literature to simulate gas-solid reactions at the scale of a single particle. An alternative approach to the numerical particle models is the law of additive reaction times presented by H.Y. Sohn in 1978. This is an approximate relationship enabling to calculate easily the time necessary to reach a given conversion in given surrounding conditions and, in its differential form, to calculate the reaction rate in the same conditions. Like particle models, it can account for different or mixed kinetic regimes. But unlike these numerical models, its simple, analytical form is a key advantage for modeling multiparticle reactors. Surprisingly, few studies were published showing its application.

We used this law for modeling a complex gas-solid reactor, the moving bed furnace converting $UO_3$ to $UF_4$. The process is presented, as well as the principal characteristics of the model, further detailed in [22-24]. In the last part of the present paper, we have examined the validity of the approach using the law of additive reaction times. We have compared its use to that of a particle model. It is shown that the law of additive reaction times predicts quite well the evolution of the conversion under different external conditions, even non-isothermal ones. Provided the characteristic reaction times are specifically calculated to precisely fit the situation to simulate, we do believe that the law of additive reaction times has a broad scope and is a powerful tool for modeling multiparticle reactors.

## Acknowledgement

The authors are very grateful to Dr. Jean Jourde, who developed the first version of the model, and thank Dr. François Nicolas for helpful discussions and continuous support. The modeling of the moving bed furnace was carried out with the support of Comurhex, a subsidiary of Areva.